\documentclass{article}
\usepackage{graphicx} 
\usepackage{authblk}
\usepackage[numbers]{natbib}
\usepackage{booktabs}
\usepackage{url} 

\title{Cancer-Net PCa-Seg: Benchmarking Deep Learning Models for Prostate Cancer Segmentation Using Synthetic Correlated Diffusion Imaging}

\author[1]{Jarett Dewbury}
\author[1]{Chi-en Amy Tai}
\author[1]{Alexander Wong}

\affil[1]{Vision and Image Processing Group, Systems Design Engineering, University of Waterloo}
\affil[ ]{\texttt{\{{jdewbury, amy.tai, alexander.wong}\}@uwaterloo.ca}}
\date{January 13, 2025}

\begin{document}

\maketitle

\begin{abstract}
Prostate cancer (PCa) is the most prevalent cancer among men in the United States, accounting for nearly 300,000 cases, 29\% of all diagnoses and 35,000 total deaths in 2024. Traditional screening methods such as prostate-specific antigen (PSA) testing and magnetic resonance imaging (MRI) have been pivotal in diagnosis, but have faced limitations in specificity and generalizability. In this paper, we explore the potential of enhancing PCa gland segmentation using a novel MRI modality called synthetic correlated diffusion imaging (CDI$^s$). We employ several state-of-the-art deep learning models, including U-Net, SegResNet, Swin UNETR, Attention U-Net, and LightM-UNet, to segment prostate glands from a 200 CDI$^s$ patient cohort. We find that SegResNet achieved superior segmentation performance with a Dice-Sørensen coefficient (DSC) of $76.68 \pm 0.8$. Notably, the Attention U-Net, while slightly less accurate (DSC $74.82 \pm 2.0$), offered a favorable balance between accuracy and computational efficiency. Our findings demonstrate the potential of deep learning models in improving prostate gland segmentation using CDI$^s$ to enhance PCa management and clinical support. 
\end{abstract}

\section{Introduction}

Prostate cancer (PCa) is the most prevalent cancer among men in the United States, accounting for nearly 300,000 cases, 29\% of all diagnoses and 35,000 total deaths in 2024~\citep{siegel2024cancer}. Although a serious disease, early detection and accurate diagnosis have helped reduce the PCa death rate by half from 1993 to 2013~\citep{AmericanCancerSocietyProstateCancer}. Traditional screening methods, such as prostate-specific antigen (PSA) testing, involve measuring the levels of PSA, a protein produced by both cancerous and non-cancerous cells, in the blood~\citep{NCI_PSATest}. Elevated PSA levels can indicate the presence of PCa, prompting additional diagnostic procedures to confirm the diagnosis, such as a prostate biopsy~\citep{NCI_PSATest}. However, PSA testing has limitations in specificity and can lead to overdiagnosis and unnecessary interventions that can be harmful for patients~\citep{loeb2014overdiagnosis}. 

In recent years, magnetic resonance imaging (MRI) has emerged as a powerful tool for prostate cancer detection. Multi-parametric MRI (mpMRI) offers detailed visualization of the prostate tissue, enabling more reliable identification and characterization of potentially clinically significant regions~\citep{stabile2020multiparametric}. The integration of deep learning techniques has further enhanced the potential of MRI-based prostate cancer diagnosis. Convolutional Neural Networks (CNNs) have shown promising results in automated prostate cancer detection and segmentation tasks, achieving statistically comparable performance metrics to those of experienced radiologists~\citep{cai2024fully}.

Despite these advancements, there are inherent limitations in traditional MRI techniques, including variability in imaging modalities and challenges in distinguishing between cancerous and non-cancerous tissues. These issues often result in poor generalization of artificial intelligence (AI) models when applied to diverse patient populations and imaging conditions, making them obsolete in practice. Several studies have explored the use of segmentation models for PCa lesion detection with a relatively low degree of success. Gunashekar et al. investigated the use of a 3D-UNet model on a 122 patient mpMRI dataset, achieving a 
Dice-Sørensen coefficient (DSC) of 0.32 when trained on 150 epochs~\citep{gunashekar2022explainable}. Similarly, Pellicer-Valero et al. demonstrated the use of a 3D Retina-UNet model across two different datasets (with 204 and 221 patients respectively), achieving a best PCa lesion DSC of 0.276~\citep{pellicer2022deep}. These results underscore the complexity of PCa lesion segmentation across considerable patient cohorts. Accurate segmentation of the prostate gland is a critical prerequisite for effective lesion detection, as it defines the anatomical context necessary for localization. Poor gland segmentation can propagate errors throughout the evaluation pipeline, leading to decreased model performance and limiting clinical utility. 

Synthetic correlated diffusion imaging (CDI$^s$) has shown superior performance in addressing the limitations of other MRI techniques by providing more distinct visualization of clinically relevant regions of interest ~\citep{wong2022synthetic}. CDI$^s$ enhances the contrast between cancerous and healthy tissue, leading to clearer delineation and reduced false positives. While AI has been applied to CDI$^s$ for breast cancer classification and demonstrated improved performance over traditional MRI modalities for grade and post-treatment response prediction~\citep{tai2023enhancing}, its application for prostate region segmentation has yet to be explored. 

In this study, we benchmark several state-of-the-art segmentation models using CDI$^s$ data to evaluate their efficacy for prostate gland segmentation in PCa patients. We aim to evaluate these models based on DSC~\citep{sorensen1948method}, inference time per patient volume, and parameter size, offering a systematic comparison of their performance and viability in a clinical setting. By identifying strengths and weaknesses in existing models, our findings can help guide future model development and selection to enhance PCa diagnostic capabilities and promote better patient outcomes. 

\section{Methodology}
\label{sec:method}
\subsection{Dataset}

\begin{figure*}[t]
  \centering
  \includegraphics[width=\linewidth]{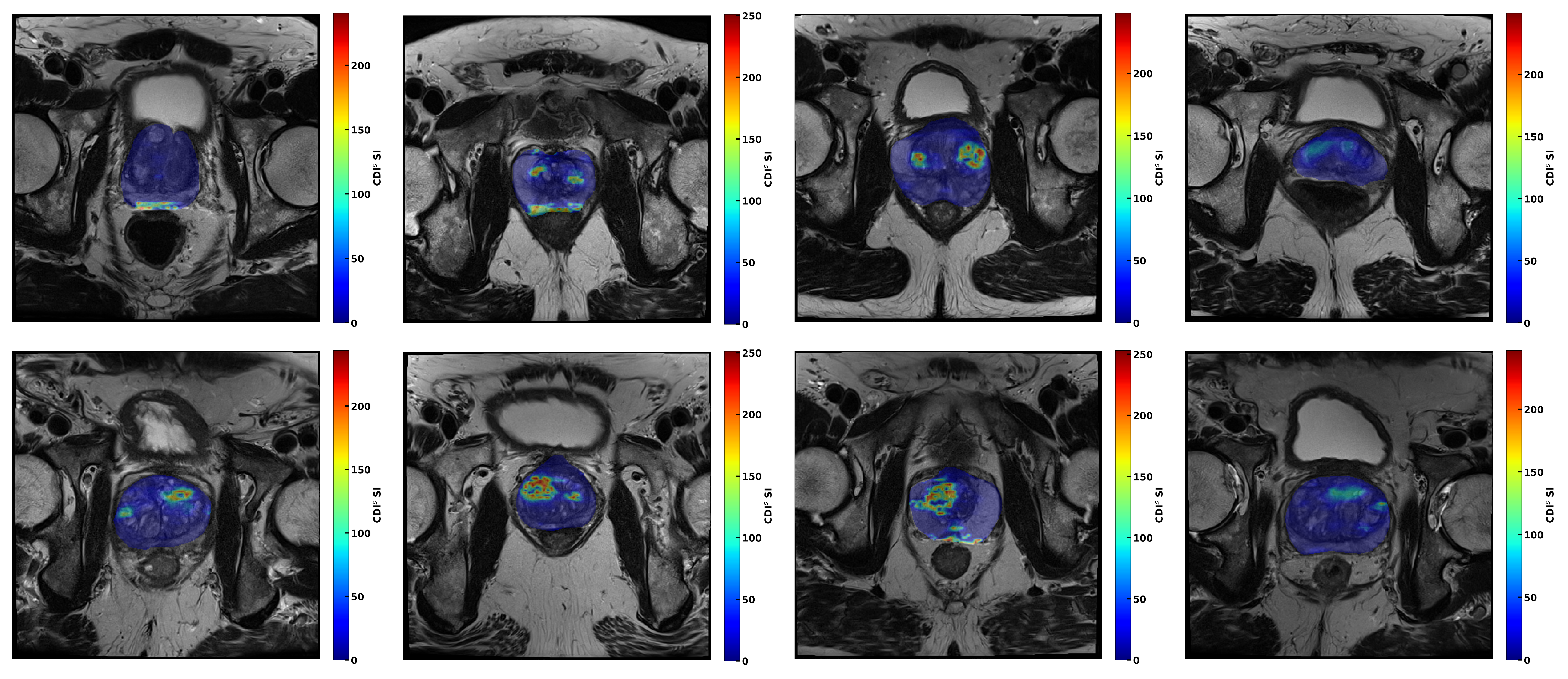}
  \caption{Visualization of sample data from the patient cohort in Cancer-Net PCa-Data dataset. T2-weighted (T2w) images are shown with CDI$^s$ prostate gland region boundary in color overlay. For the purposes of our paper, we consider both clinically significant (csPCa) and clinically insignificant (insPCa) prostate cancer segmentation.}
  \label{fig:figure_1}
\end{figure*}

\begin{figure*}[t]
  \centering
  \includegraphics[width=\linewidth]{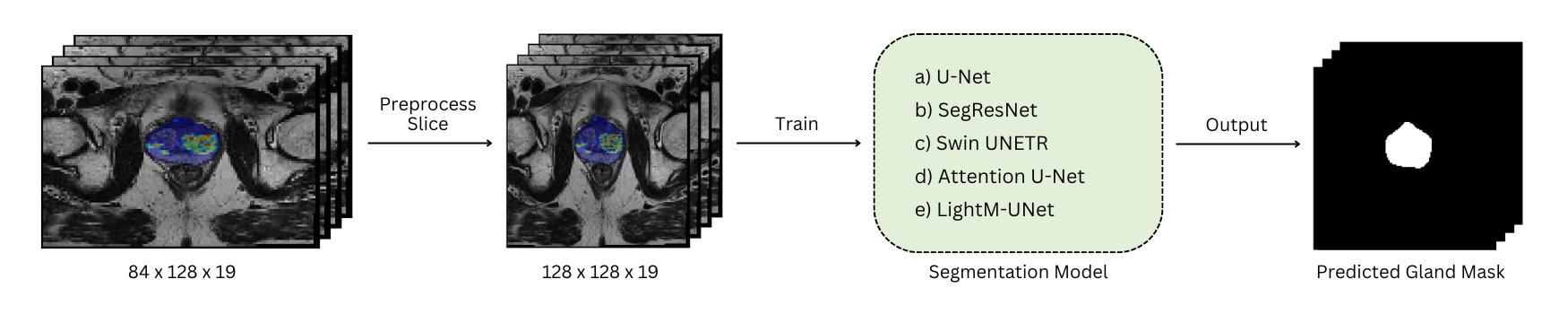}
  \caption{Prostate cancer segmentation pipeline using Cancer-Net PCa-Data. The workflow includes (left to right): input CDI$^s$ volume slice, preprocessing, segmentation across five different deep learning models, and the final predicted prostate gland mask output.}
  \label{fig:figure_2}
\end{figure*}

This study leveraged the Cancer-Net PCa-Data dataset~\citep{gunraj2023cancer}, a novel and publicly available benchmark dataset for PCa research. This data was derived from the SPIE-AAPM-NCI PROSTATEx Challenges, PROSTATEx\_masks, and The Cancer Imaging Archive (TCIA) datasets~\citep{Litjens2014,Litjens2017,Cuocolo2021,Clark2013}. The dataset contains volumetric CDI$^s$ imaging data for 200 patients along with their associated prostate gland and tumor lesion mask (see Figure~\ref{fig:figure_1} for sample patient data). 

\subsection{Data Preparation}

A training, validation, and test set split of 70\%, 15\%, and 15\% was used respectively. The data was split at the patient level, ensuring the entire set of slices from a single patient volume were assigned to the same data split to prevent data leakage. As illustrated in Figure~\ref{fig:figure_2}, the preprocessing pipeline involved standardizing the image volumes and corresponding masks using min-max normalization and resizing to 128x128x19, with a batch size of 16. This resulted in a distribution of 2736 training slices, 592 validation slices, and 592 test slices.

\subsection{Training}

We evaluated the baseline performance on the dataset using commonly used model architectures in medical image segmentation tasks including U-Net~\citep{ronneberger2015unetconvolutionalnetworksbiomedical}, SegResNet~\citep{myronenko20183dmribraintumor}, Swin UNETR~\citep{hatamizadeh2022swinunetrswintransformers}, Attention U-Net~\citep{oktay2018attentionunetlearninglook}, and LightM-UNet~\citep{liao2024lightm}. All models were implemented using the MONAI framework~\citep{MONAI2024}, except for LightM-UNet, which utilized the PyTorch implementation provided in their paper~\citep{liao2024lightm}. Each model was trained for 200 epochs using Binary Cross Entropy with Logits Loss as the primary loss metric. A learning rate of 0.001 was used for all models except LightM-UNet, which was trained with a learning rate of 0.0001. This adjustment significantly improved the performance of LightM-UNet on the dataset. A StepLR scheduler was employed with a step size of 0.2 adjusting the learning rate every 50 epochs. The vanilla Adam optimizer was utilized across all experiments, and no pre-training was conducted on any of the models. Model performance was primarily evaluated using DSC, with the best validation DSC determining the model weights used for testing. To ensure robustness of model performance, each model was trained and evaluated across five different data fold seeds 79, 46, 39, 76, and 42.  

\section{Results}

\begin{table*}[b]
    \centering
    \caption{Experimental results of DSC on the testing set (standard deviation indicated after $\pm$), inference time in seconds for a single patient volume, and number of parameters in millions reported across the evaluated models.}
    \label{tab:MS_main_result}
    \begin{tabular}{l c c c}
    \toprule
    \multicolumn{1}{c}{\textbf{Model}} & \multicolumn{1}{c}{\textbf{DSC}} & \multicolumn{1}{c}{\textbf{Inference Time (s)}} & \multicolumn{1}{c}{\textbf{Params (M)}}\\
    \midrule
    U-Net~\citep{ronneberger2015unetconvolutionalnetworksbiomedical} & $71.35 \pm 1.5$ & $0.0013$ & $1.62$\\
    SegResNet~\citep{myronenko20183dmribraintumor} & $76.68 \pm 0.8$ & $0.0016$ & $6.29$\\
    Swin UNETR~\citep{hatamizadeh2022swinunetrswintransformers} & $73.46 \pm 1.8$ & $0.0021$ & $6.30$ \\
    Attention U-Net ~\citep{oktay2018attentionunetlearninglook} & $74.82 \pm 2.0$ & $0.0015$ & $1.99$ \\
    LightM-UNet~\citep{liao2024lightm} & $71.38 \pm 1.5$ & $0.0016$ & $2.30$ \\
    \bottomrule
    \end{tabular}
\end{table*}

Table~\ref{tab:MS_main_result} presents a comparative analysis of the evaluated segmentation models and performance on the testing set across DSC, inference time per patient volume, and total parameter count. 

\label{sec:results}
Model results are presented as mean $\pm$ standard deviation (n=5), where n denotes the total number of independent data folds evaluated on. The THOP library was used to calculate the total number of parameters for each model~\citep{ligengzhupytorch}. 
 
The baseline U-Net achieved a DSC of $71.35 \pm 1.5$ with a significantly smaller parameter count of 1.62 million and the fastest inference time of 0.0013 seconds. Its lightweight architecture and fast inference time make the U-Net particularly suitable for applications where real-time processing speed and resource efficiency is critical over the cost of a marginal improvement in segmentation accuracy. 

SegResNet achieved the highest test set DSC of $76.68 \pm 0.8$, indicating superior prostate tumor segmentation capabilities on the Cancer-Net PCa-Data dataset. This suggests that SegResNet is particularly effective in capturing the prostate gland boundaries, making it a strong candidate for precise prostate region segmentation applications. However, SegResNet's superior performance comes at a tradeoff of increased computational demands, with the second largest number of parameters and inference time amongst the architectures of 6.29 million and 0.0016 seconds per patient volume, respectively. This suggests that while SegResNet offers optimal prostate gland segmentation performance out of the evaluated models, its computational demands may limit its applicability in high-throughput clinical scenarios. 

Swin UNETR, despite its transformer-based architecture, demonstrated moderate segmentation performance improvements over the U-Net with a DSC of $73.46 \pm 1.8$ relative to its higher parameter count (6.30 million) and inference time (0.0021 seconds). This limits its applicability in real-time or resource constrained scenarios for PCa segmentation. 

Attention U-Net exhibited a notable improvement over the baseline U-Net, with the second highest DSC of $74.82 \pm 2.0$ and a slight increase in parameters (1.99 million) and inference time (0.0015 seconds). This suggests that the attention mechanism effectively enhances feature selection for identifying the prostate gland without substantially increasing complexity. 

The LightM-UNet's performance was comparable to the U-Net with a DSC of $71.38 \pm 1.5$, 2.30 million total parameters, and inference time of 0.0016 seconds per patient volume. Despite its architectural innovations, its marginal performance difference relative to the baseline U-Net suggests that the design advantages of the LightM-UNet may not be fully leveraged by the training pipeline. 

\section{Conclusion}
\label{sec:concl}
In this paper, we investigate the performance of various deep learning models to enhance prostate gland segmentation for patients with prostate cancer (PCa) using a novel MRI modality called synthetic correlated diffusion imaging (CDI$^s$). SegResNet demonstrated superior segmentation performance with a Dice-Sørensen coefficient (DSC) of $76.68 \pm 0.8$, showcasing its potential for precise prostate gland delineation. However, its higher computational demands may limit its applicability in real-time clinical scenarios. Notably, Attention U-Net emerged as a promising compromise between performance and efficiency. With the second highest DSC of $74.82 \pm 2.0$ and relatively low computational requirements, it represents a viable option for clinical implementation. Given the promising results of the deep learning models on the CDI$^s$ dataset, future work involves exploring additional model architectures that could further optimize segmentation performance and computational efficiency. The transferability of these models on additional CDI$^s$ patient cohorts should also be explored to assess the robustness of the models in clinical practice across diverse patient populations.

\newpage

{
\small

\bibliographystyle{unsrt}
\bibliography{refs}

\begin{thebibliography}{10}

\bibitem{siegel2024cancer}
Rebecca~L Siegel, Angela~N Giaquinto, and Ahmedin Jemal.
\newblock Cancer statistics, 2024.
\newblock {\em CA: a cancer journal for clinicians}, 74(1), 2024.

\bibitem{AmericanCancerSocietyProstateCancer}
American~Cancer Society.
\newblock Prostate cancer facts.
\newblock \url{https://www.cancer.org/cancer/types/prostate-cancer/about/key-statistics.html}, 2024.
\newblock Accessed: 2024-08-30.

\bibitem{NCI_PSATest}
National~Cancer Institute.
\newblock What is the psa test?
\newblock \url{https://www.cancer.gov/types/prostate/psa-fact-sheet#:~:text=the%20PSA%20test%3F-,What%20is%20the%20PSA%20test%3F,to%20a%20laboratory%20for%20analysis.}, 2024.
\newblock Accessed: 2024-08-30.

\bibitem{loeb2014overdiagnosis}
Stacy Loeb, Marc~A Bjurlin, Joseph Nicholson, Teuvo~L Tammela, David~F Penson, H~Ballentine Carter, Peter Carroll, and Ruth Etzioni.
\newblock Overdiagnosis and overtreatment of prostate cancer.
\newblock {\em European urology}, 65(6):1046--1055, 2014.

\bibitem{stabile2020multiparametric}
Armando Stabile, Francesco Giganti, Andrew~B Rosenkrantz, Samir~S Taneja, Geert Villeirs, Inderbir~S Gill, Clare Allen, Mark Emberton, Caroline~M Moore, and Veeru Kasivisvanathan.
\newblock Multiparametric mri for prostate cancer diagnosis: current status and future directions.
\newblock {\em Nature reviews urology}, 17(1):41--61, 2020.

\bibitem{cai2024fully}
Jason~C Cai, Hirotsugu Nakai, Shiba Kuanar, Adam~T Froemming, Candice~W Bolan, Akira Kawashima, Hiroaki Takahashi, Lance~A Mynderse, Chandler~D Dora, Mitchell~R Humphreys, et~al.
\newblock Fully automated deep learning model to detect clinically significant prostate cancer at mri.
\newblock {\em Radiology}, 312(2):e232635, 2024.

\bibitem{gunashekar2022explainable}
Deepa~Darshini Gunashekar, Lars Bielak, Leonard H{\"a}gele, Benedict Oerther, Matthias Benndorf, Anca-L Grosu, Thomas Brox, Constantinos Zamboglou, and Michael Bock.
\newblock Explainable ai for cnn-based prostate tumor segmentation in multi-parametric mri correlated to whole mount histopathology.
\newblock {\em Radiation Oncology}, 17(1):65, 2022.

\bibitem{pellicer2022deep}
Oscar~J Pellicer-Valero, Jose~L Marenco~Jimenez, Victor Gonzalez-Perez, Juan~Luis Casanova Ramon-Borja, Isabel Mart{\'\i}n~Garc{\'\i}a, Maria Barrios~Benito, Paula Pelechano~Gomez, Jos{\'e} Rubio-Briones, Mar{\'\i}a~Jos{\'e} Rup{\'e}rez, and Jos{\'e}~D Mart{\'\i}n-Guerrero.
\newblock Deep learning for fully automatic detection, segmentation, and gleason grade estimation of prostate cancer in multiparametric magnetic resonance images.
\newblock {\em Scientific reports}, 12(1):2975, 2022.

\bibitem{wong2022synthetic}
Alexander Wong, Hayden Gunraj, Vignesh Sivan, and Masoom~A Haider.
\newblock Synthetic correlated diffusion imaging hyperintensity delineates clinically significant prostate cancer.
\newblock {\em Scientific Reports}, 12(1):3376, 2022.

\bibitem{tai2023enhancing}
Chi-en~Amy Tai, Hayden Gunraj, Nedim Hodzic, Nic Flanagan, Ali Sabri, and Alexander Wong.
\newblock Enhancing clinical support for breast cancer with deep learning models using synthetic correlated diffusion imaging.
\newblock In {\em International Workshop on Applications of Medical AI}, pages 83--93. Springer, 2023.

\bibitem{sorensen1948method}
Thorvald Sorensen.
\newblock A method of establishing groups of equal amplitude in plant sociology based on similarity of species content and its application to analyses of the vegetation on danish commons.
\newblock {\em Biologiske skrifter}, 5:1--34, 1948.

\bibitem{gunraj2023cancer}
Hayden Gunraj, Chi en~Amy~Tai, and Alexander Wong.
\newblock Cancer-net pca-data: An open-source benchmark dataset for prostate cancer clinical decision support using synthetic correlated diffusion imaging data.
\newblock {\em NeurIPS Workshops}, 2023.

\bibitem{Litjens2014}
Geert Litjens, Oscar Debats, Jelle Barentsz, Nico Karssemeijer, and Henkjan Huisman.
\newblock Computer-aided detection of prostate cancer in mri.
\newblock {\em IEEE Transactions on Medical Imaging}, 33(5):1083--1092, 2014.

\bibitem{Litjens2017}
Geert Litjens, Oscar Debats, Jelle Barentsz, Nico Karssemeijer, and Henkjan Huisman.
\newblock Prostatex challenge data [data set], 2017.

\bibitem{Cuocolo2021}
Renato Cuocolo, Arnaldo Stanzione, Anna Castaldo, Davide~Raffaele {De Lucia}, and Massimo Imbriaco.
\newblock Quality control and whole-gland, zonal and lesion annotations for the prostatex challenge public dataset.
\newblock {\em European Journal of Radiology}, 138:109647, 2021.

\bibitem{Clark2013}
Kenneth Clark, Bruce Vendt, Kirk Smith, John Freymann, Justin Kirby, Paul Koppel, Stephen Moore, Stanley Phillips, David Maffitt, Michael Pringle, Lawrence Tarbox, and Fred Prior.
\newblock The cancer imaging archive (tcia): Maintaining and operating a public information repository.
\newblock {\em Journal of Digital Imaging}, 26(6):1045--1057, 2013.

\bibitem{ronneberger2015unetconvolutionalnetworksbiomedical}
Olaf Ronneberger, Philipp Fischer, and Thomas Brox.
\newblock U-net: Convolutional networks for biomedical image segmentation.
\newblock In {\em Medical image computing and computer-assisted intervention--MICCAI 2015: 18th international conference, Munich, Germany, October 5-9, 2015, proceedings, part III 18}, pages 234--241. Springer, 2015.

\bibitem{myronenko20183dmribraintumor}
Andriy Myronenko.
\newblock 3d mri brain tumor segmentation using autoencoder regularization.
\newblock In {\em Brainlesion: Glioma, Multiple Sclerosis, Stroke and Traumatic Brain Injuries: 4th International Workshop, BrainLes 2018, Held in Conjunction with MICCAI 2018, Granada, Spain, September 16, 2018, Revised Selected Papers, Part II 4}, pages 311--320. Springer, 2019.

\bibitem{hatamizadeh2022swinunetrswintransformers}
Ali Hatamizadeh, Vishwesh Nath, Yucheng Tang, Dong Yang, Holger~R Roth, and Daguang Xu.
\newblock Swin unetr: Swin transformers for semantic segmentation of brain tumors in mri images.
\newblock In {\em International MICCAI brainlesion workshop}, pages 272--284. Springer, 2021.

\bibitem{oktay2018attentionunetlearninglook}
Ozan Oktay, Jo~Schlemper, Loic~Le Folgoc, Matthew Lee, Mattias Heinrich, Kazunari Misawa, Kensaku Mori, Steven McDonagh, Nils~Y Hammerla, Bernhard Kainz, et~al.
\newblock Attention u-net: Learning where to look for the pancreas.
\newblock {\em arXiv preprint arXiv:1804.03999}, 2018.

\bibitem{liao2024lightm}
Weibin Liao, Yinghao Zhu, Xinyuan Wang, Chengwei Pan, Yasha Wang, and Liantao Ma.
\newblock Lightm-unet: Mamba assists in lightweight unet for medical image segmentation.
\newblock {\em arXiv preprint arXiv:2403.05246}, 2024.

\bibitem{MONAI2024}
MONAI.
\newblock Monai: Medical open network for ai.
\newblock \url{https://monai.io/}, 2024.
\newblock Accessed: 2024-08-30.

\bibitem{ligengzhupytorch}
Ligeng Zhu.
\newblock Thop: Pytorch-opcounter.
\newblock \url{https://github.com/Lyken17/pytorch-OpCounter}, 2022.
\newblock Accessed: 2024-07-21.

\end{thebibliography}
}

\end{document}